\RequirePackage{fix-cm} 
\documentclass[a4paper, twoside, reqno, 12pt, dvips]{amsart}

\usepackage{fixltx2e}     

\usepackage{amssymb}
\usepackage{amsmath}
\usepackage{mathrsfs, dsfont}

\usepackage{a4wide}

\usepackage{indentfirst}
\usepackage{graphicx}
\usepackage{psfrag}

\usepackage[usenames,dvipsnames]{pstricks}
\usepackage{epsfig}
\usepackage{pst-grad} 
\usepackage{pst-plot} 
\usepackage{infix-RPN}

\usepackage{subfigure}

\usepackage{acronym}
\usepackage{longtable}

\usepackage[english]{babel}
\usepackage[latin1]{inputenc}

\usepackage{color}
\definecolor{oneblue}{rgb}{0,0.0,0.75}
\usepackage[colorlinks,
            urlcolor=oneblue,
            linkcolor=oneblue,
            citecolor=oneblue,
            bookmarksopen=false,
            pagebackref]{hyperref}

\numberwithin{equation}{section}

\def\R{\mathbb{R}}

\begin{document}

\title[Local Runup Amplification]{Local Runup Amplification By Resonant Wave Interactions}

\author[T. Stefanakis]{Themistoklis Stefanakis}
\address{CMLA, ENS Cachan, 61 avenue du Pr\'esident Wilson, 94235 Cachan Cedex, France and School of Mathematical Sciences, University College Dublin, Belfield, Dublin 4, Ireland}
\email{Themistoklis.Stefanakis@cmla.ens-cachan.fr}

\author[F. Dias]{Fr\'ed\'eric Dias$^*$}
\address{CMLA, ENS Cachan, 61 avenue du Pr\'esident Wilson, 94235 Cachan Cedex, France and School of Mathematical Sciences, University College Dublin, Belfield, Dublin 4, Ireland}
\email{Frederic.Dias@ucd.ie}
\thanks{$^*$ Corresponding author}

\author[D. Dutykh]{Denys Dutykh}
\address{LAMA, UMR 5127 CNRS, Universit\'e de Savoie, Campus Scientifique, 73376 Le Bourget-du-Lac Cedex, France}
\email{Denys.Dutykh@univ-savoie.fr}
\urladdr{http://www.lama.univ-savoie.fr/~dutykh/}

\begin{abstract}
Until now the analysis of long wave runup on a plane beach has been focused on finding its maximum value, failing to capture the existence of resonant regimes. One-dimensional numerical simulations in the framework of the Nonlinear Shallow Water Equations (NSWE) are used to investigate the Boundary Value Problem (BVP) for plane and non-trivial beaches. Monochromatic waves, as well as virtual wave-gage recordings from real tsunami simulations, are used as forcing conditions to the BVP. Resonant phenomena between the incident wavelength and the beach slope are found to occur, which result in enhanced runup of non-leading waves.  The evolution of energy reveals the existence of a quasi-periodic state for the case of sinusoidal waves, the energy level of which, as well as the time required to reach that state, depend on the incident wavelength for a given beach slope. Dispersion is found to slightly reduce the value of maximum runup, but not to change the overall picture. Runup amplification occurs for both leading elevation and depression waves.
\end{abstract}


\maketitle

\tableofcontents

\section{Introduction}

Despite mathematical difficulties, wave runup, which is the maximum vertical extent of wave uprush on a beach above still water level \cite{Sorensen1997},  has been extensively studied during the last fifty years. Progress was first made to the one-dimensional long wave problem. From the 1950's until 1990 several major contributions were made to the initial value problem (IVP) of long wave runup \cite{CG58, Keller1964, Carrier1966, Synolakis1986, Synolakis1987}, mainly through the use of the \cite{CG58} (CG) transformation, that allows the reduction of the two NSWE into a single linear equation. After the two 1992 tsunamis (Nicaragua and Flores Island), measurements suggested that the shoreline receded before inundation took place, an observation that lead \cite{TS94} to propose a new N-shaped wave profile as a leading wave model. Recently, a more geophysically  relevant N-wave model was derived and the resulting runup on a plane beach was computed \cite{Madsen2010}. Apart from the plane beach geometry, wave evolution and runup have also been addressed for piecewise linear topographies \cite{Kanoglu1998}. Expressions for long wave runup that are independent of the initial waveform were derived by \cite{Didenkulova2008}. All the above results dealt with the IVP. \cite{Antuono2007} solved the BVP for the NSWE, using the CG \cite{CG58} transformation and applied a perturbation approach by assuming small incoming waves at the seaward boundary. Later, the same authors \cite{Antuono2010} solved the BVP in physical space without use of the CG \cite{CG58} transformation.

Concerning the two-dimensional problem the sole analytical solution was derived by \cite{Brocchini1996} who used a transformation to relate the longshore coordinate to the time variable. This operation allowed them to use an expression for the horizontal velocity that reduced the dimensions and transformed their problem into the already solved one-dimensional canonical problem. However their solution is only valid for mild angles of incidence.

Almost all of the aforementioned studies focus on the value of the maximum runup. Nevertheless, extreme runup values measured by field studies, like during the 17 July 2006 Java event \cite{Fritz2007}, cannot be explained by the existing theory. Furthermore, in some cases, on the aftermath of a tsunami catastrophe it has been reported that it was not the first tsunami wave that caused the maximum damage. In order to explain this phenomenon, scientists assume that the amplified maximum runup values of non-leading tsunami waves are due to reflection and refraction effects from nearshore topographic features \cite{Neetu2011}. It is found that changes in bathymetry (i.e. underwater topography) may result in wave resonance \cite{Miles1967, Kajiura1977, Agnon1988, Grataloup2003}. In the present study, with the use of one-dimensional numerical simulations we attempt to elucidate the runup amplification by non-leading long waves.

\section{Discussion of obtained results}

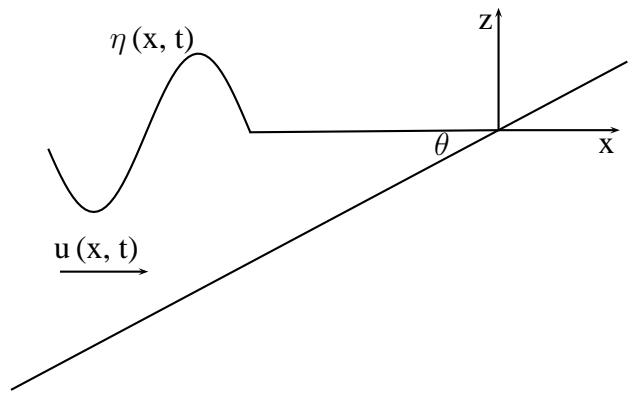
\begin{figure}
\begin{center}
\scalebox{0.7} 
{
\begin{pspicture}(0,-3.64)(11.6,3.64)
\psline[linewidth=0.04cm](0.0,-3.62)(11.58,2.6)
\psline[linewidth=0.04cm](4.48,1.26)(9.18,1.3)
\infixtoRPN{-1.5*sin(1.6*(x-4.5)*180/3.1415)+1.25}
\psplot[linewidth=0.04]
     {0.7}{4.5}{\RPN}
\psline[linewidth=0.04cm,arrowsize=0.05291667cm 2.0,arrowlength=1.4,arrowinset=0.4]{->}(9.16,1.3)(9.16,3.62)
\psline[linewidth=0.04cm,arrowsize=0.05291667cm 2.0,arrowlength=1.4,arrowinset=0.4]{->}(9.16,1.3)(11.44,1.3)
\usefont{T1}{ptm}{m}{n}
\rput(8.92,3.365){\Large z}
\usefont{T1}{ptm}{m}{n}
\rput(11.19,1.0){\Large x}
\usefont{T1}{ptm}{m}{n}
\rput(8.1,1.05){\Large $\theta$}
\psline[linewidth=0.04cm,arrowsize=0.05291667cm 2.0,arrowlength=1.4,arrowinset=0.4]{->}(0.92,-1.38)(2.58,-1.38)
\usefont{T1}{ptm}{m}{n}
\rput(1.6,-0.975){\Large u\,(x, t)}
\usefont{T1}{ptm}{m}{n}
\rput(2.65,3.0){\Large $\eta$\,(x, t)}
\end{pspicture} 
}
\end{center}
\caption{Geometry of the runup problem}
\label{fig:geometry}
\end{figure}

The maximum wave runup for the geometry of Fig. \ref{fig:geometry} was first studied for three different beach slopes, namely $\tan\theta = 0.13\, ; 0.26\, ; 0.3\, ,$ using incident monochromatic waves at $x=-L$ of the form $\eta(-L,t) = \pm \eta_0 \  \sin(\omega t), \   \omega/\sqrt{g \tan{\theta} / L} \in (0,\, 6.29)$.

\begin{figure}
\centering
\includegraphics[width=0.9\textwidth]{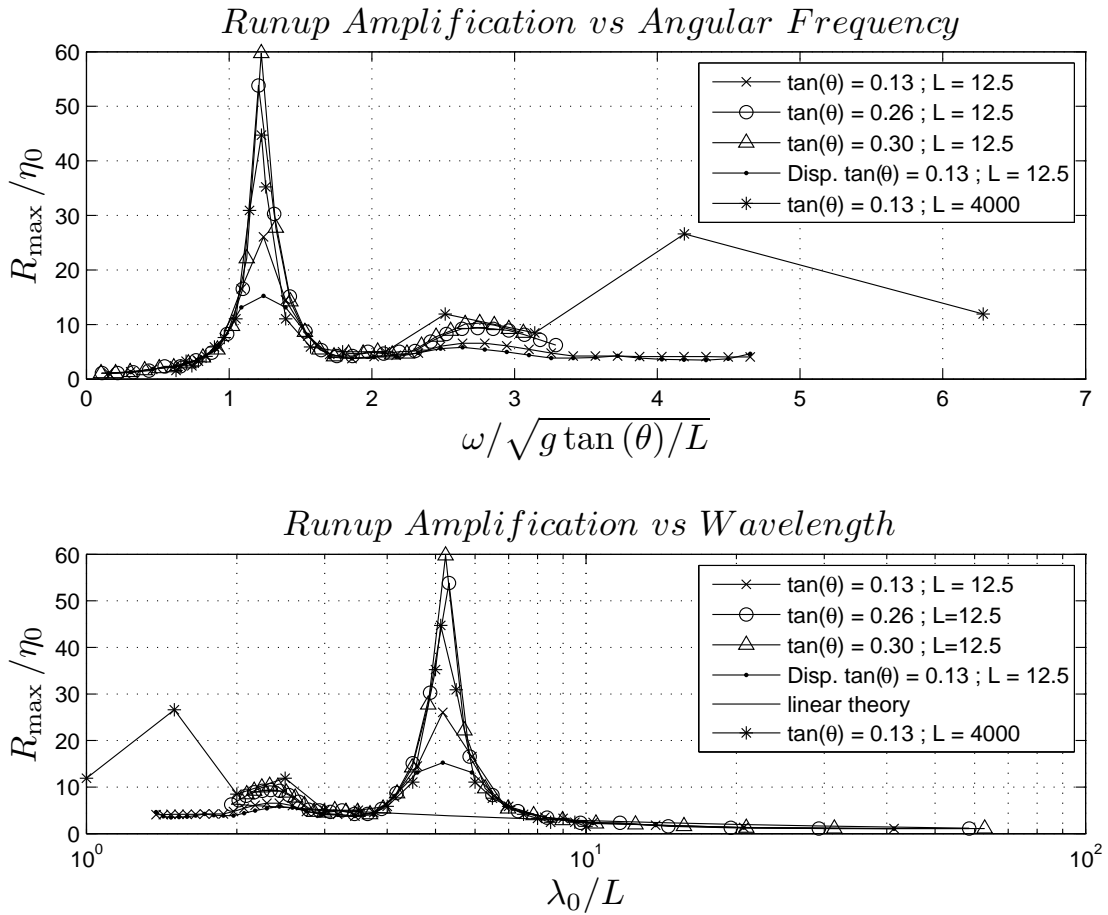}
\caption{Maximum runup amplification ratio as a function of non-dimensional angular frequency (top) and non-dimensional wavelength (bottom) for two beach lengths, namely $L=12.5$ m and $4000$ m.}
\label{fig:sin_ampli}
\end{figure}

\begin{figure}
\centering
\includegraphics[width=0.7\textwidth]{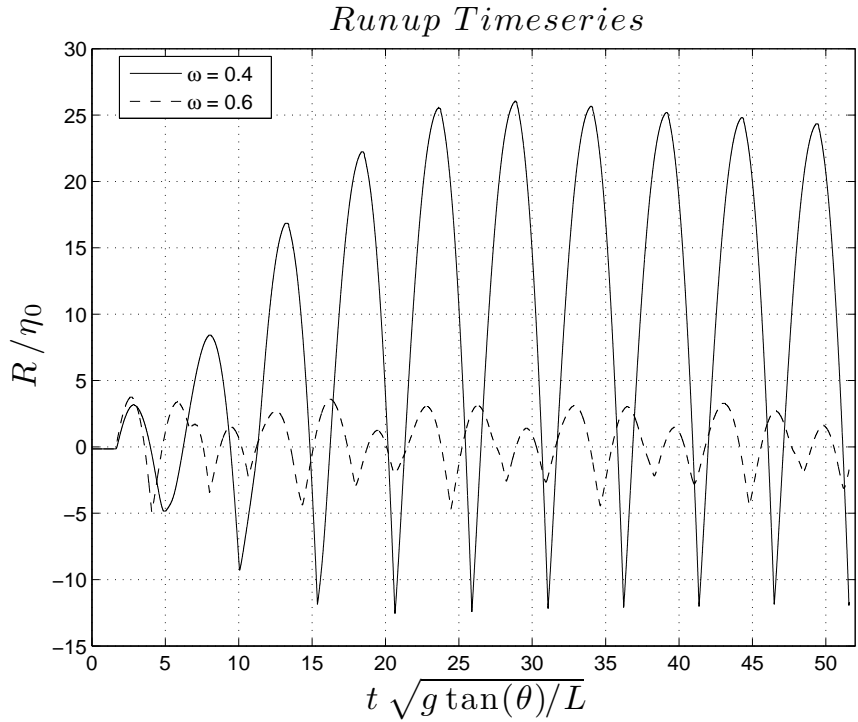}
\caption{Runup timeseries for two different angular frequencies: $\omega = 0.4 \,$ s$^{-1}$ which is the resonant frequency for $\tan\theta = 0.13$ and $\omega = 0.6 \,$ s$^{-1}$, which is a non-resonant frequency for the same slope ($L = 12.5$ m).}
\label{fig:shore_elev}
\end{figure}

The maximum runup for a given beach slope was found to depend on the incident wavelength (Fig. \ref{fig:sin_ampli}). For all three slopes, the maximum runup is highest when the non-dimensional wavelength $\lambda_0 /L \approx 5.1$ (i.e. at $L = 100\,$ m offshore the resonant wavelength is $\lambda_0 = 510\,$ m), where $\lambda_0 = 2\pi \sqrt{gL \tan\theta}/\omega$ is the wavelength of the incident wave. For increasing slope, the maximum runup also increases and reaches an amplification factor $R_{\max}/\eta_0 =  59.76$ when $\tan\theta = 0.3$, which clearly is extremely high. Increasing the beach length leads to higher resonant maximum runup values, as well as a secondary resonant regime at $\lambda_0 / L = 1.5$. Amplification is seen for both leading elevation and depression waves. Adding dispersion to the system \cite{Dutykh2010} only results in a reduction of the maximum runup value at the resonant frequencies, without qualitatively changing the overall picture. The aforementioned values of maximum runup were not achieved by the first incident wave, as is the case for $\lambda_0 /L > 10$, but by subsequent ones (Fig. \ref{fig:shore_elev}), thus signifying the existence of some resonant phenomena, the controlling parameters of which are the incident wavelength and the beach slope. Enhanced but not as extreme runup is also present for wavelengths which are approximately half the resonant ones, an observation that strengthens the assumption that the harmonics play an important role on the runup. The existence of resonant regimes is not predicted by linear theory \cite{Pelinovsky1992}, according to which $R_{\max}/\eta_0 = 2\pi \sqrt{{2L}/{\lambda_0}}$ (Fig. \ref{fig:sin_ampli}). However, the theory is in close agreement with the computed results in the absence of resonance.

Figure \ref{fig:shore_elev} also shows that waves with both resonant and non-resonant frequencies reach a quasi-periodic state of equilibrium, which is reached faster when the frequency is non-resonant. A key difference  is the existence of a single peak/trough (runup/rundown) at the quasi-periodic state of the resonant regime while the non-resonant frequencies show multiple peaks/troughs in their quasi-periodic states. This is indicative of the importance of the synchronization between the incident and reflective waves on the runup and rundown.

Next we describe this novel resonant mechanism in terms of energy. The potential and kinetic energy of the wave are respectively \cite{Dutykh2009b}: $E_K = \frac{1}{2} \rho \int_{\mathbb{R}} \int_{-h}^{\eta} u^2 dx dz$ \ and \ $E_P = \frac{1}{2} \rho g \int_{\mathbb{R}}  \eta^2 dx \,. $ The kinetic energy can be reformulated in terms of the total flow depth $H=\eta+h$ as $E_K = \frac{1}{2} \rho \int_\R H u^2 dx$ .

\begin{figure}
\centering
\includegraphics[width=0.7\textwidth]{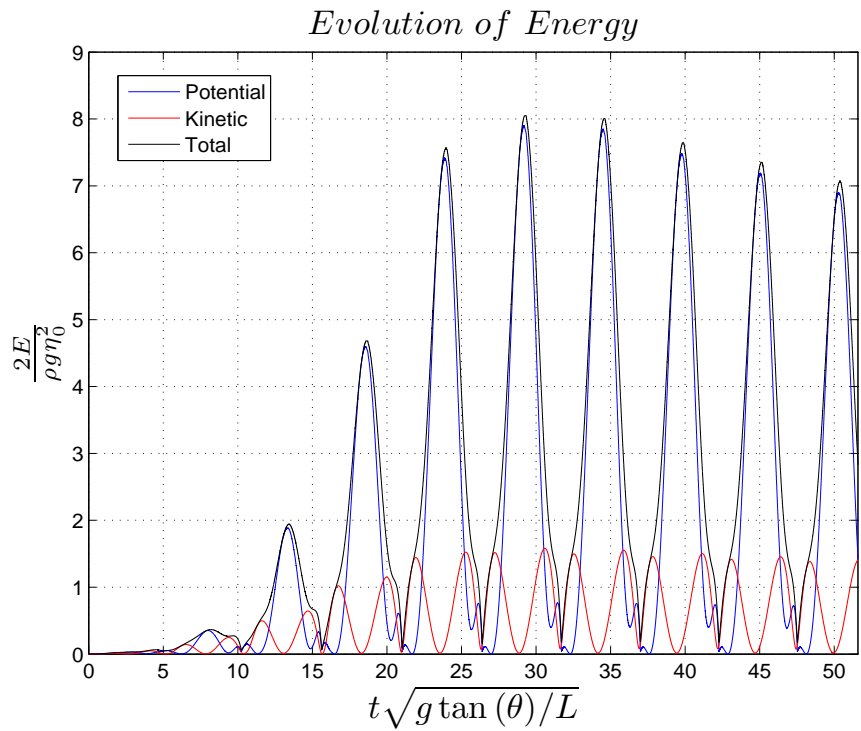}
\caption{Energy evolution timeseries for the resonant frequency $\omega = 0.4\, s^{-1}$ for $\tan\theta = 0.13$, $L = 12.5$ m.}
\label{fig:sin_engy}
\end{figure}

The evolution of the energy for the resonant frequency ($\omega = 0.4\, $s$^{-1}$) when $\tan\theta = 0.13$ is shown in Fig. \ref{fig:sin_engy}. One can see that both the maximum potential and kinetic energies increase with time until the quasi-periodic state is reached ($t \sqrt{g\tan\theta /L} \approx 30$). The potential energy takes its maximum value at the instance of the maximum runup, when the kinetic energy is minimum. However, it is obvious that the maximum potential energy is approximately five times larger than the maximum kinetic energy. The impressive oscillations in the total energy are due to the  large changes of the portion of the computational domain covered by water during runup and rundown, which actually affects the limits of integration in the energy equations.

\begin{figure}
\begin{center}
\begin{minipage}{2.2in}
{\includegraphics[height=2.0in]{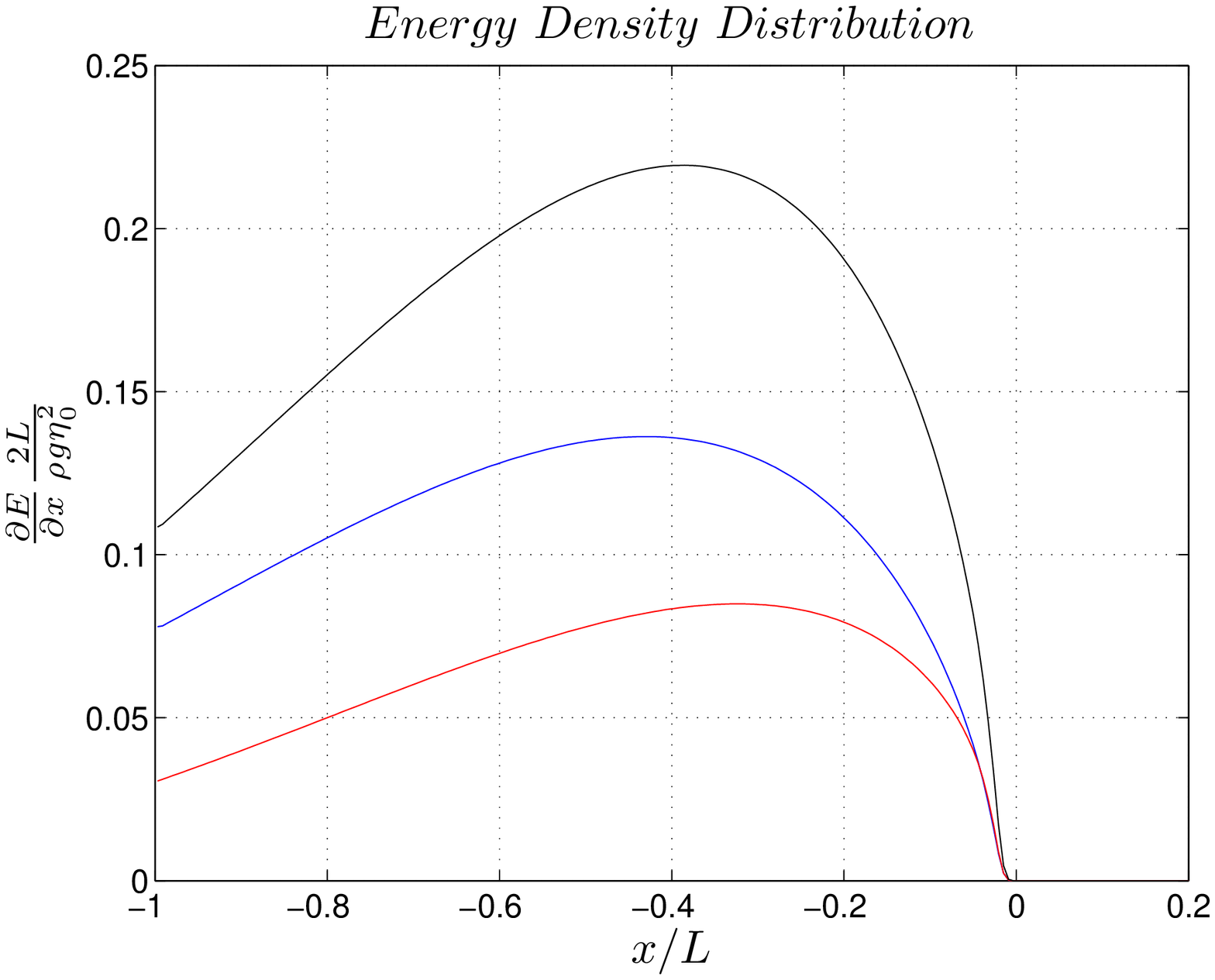}}
{\small (a)}
\end{minipage}
\hspace{0.2in}
\begin{minipage}{2.1in}
{\includegraphics[height=2.0in]{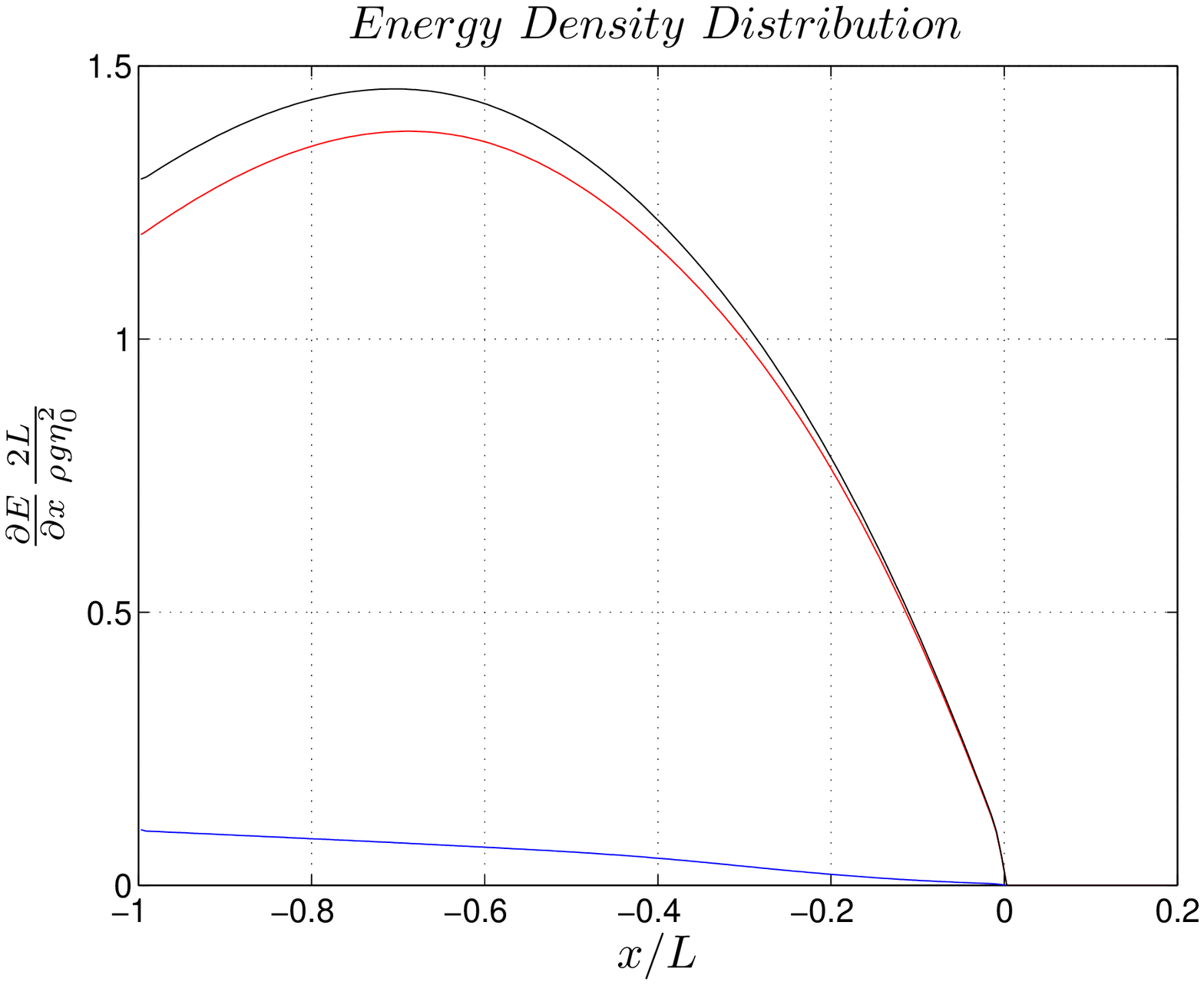}}
{\small (b)}
\end{minipage}
\begin{minipage}{2.2in}
{\includegraphics[height=2.0in]{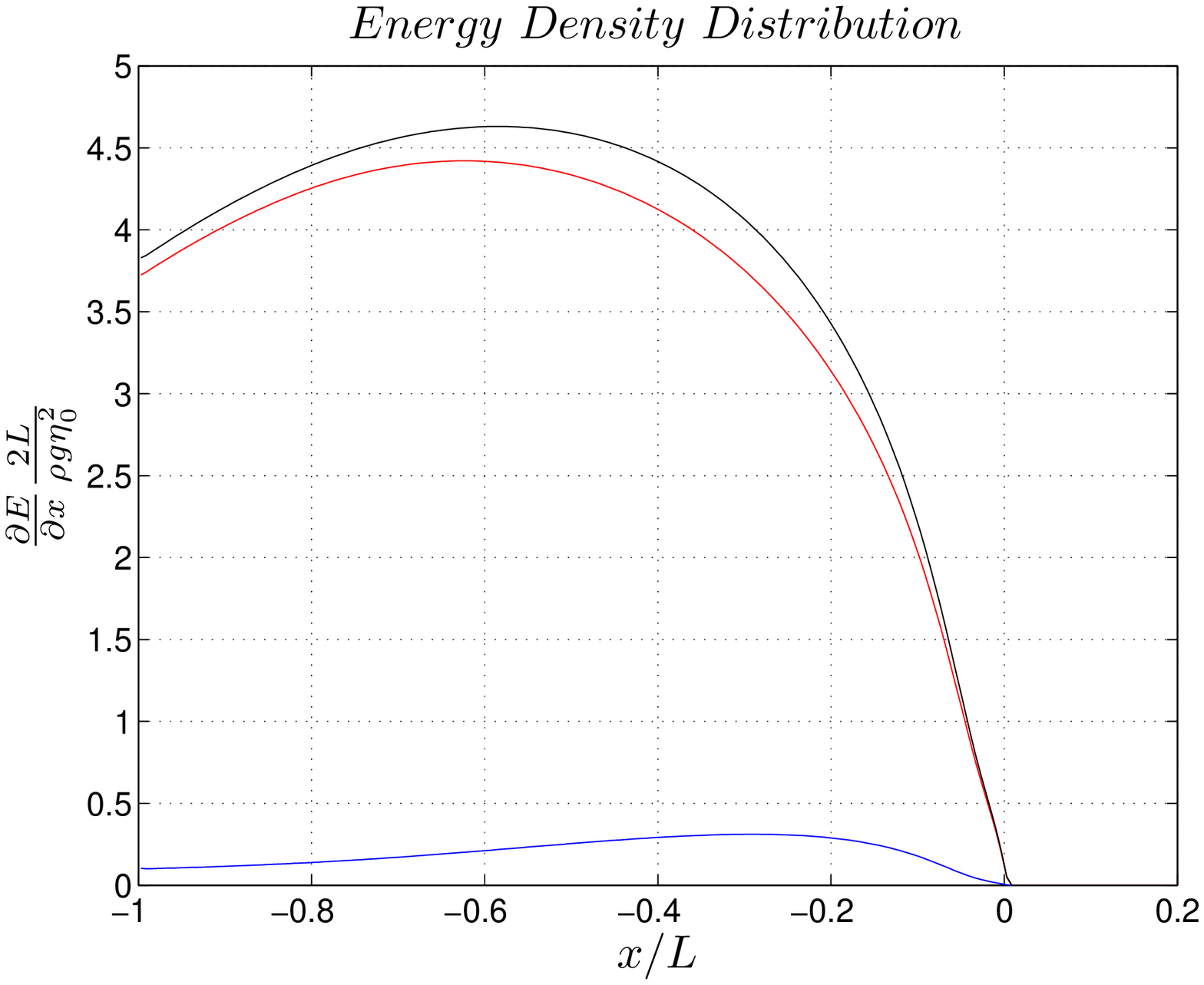}}
{\small (c)}
\end{minipage}
\hspace{0.2in}
\begin{minipage}{2.1in}
{\includegraphics[height=2.0in]{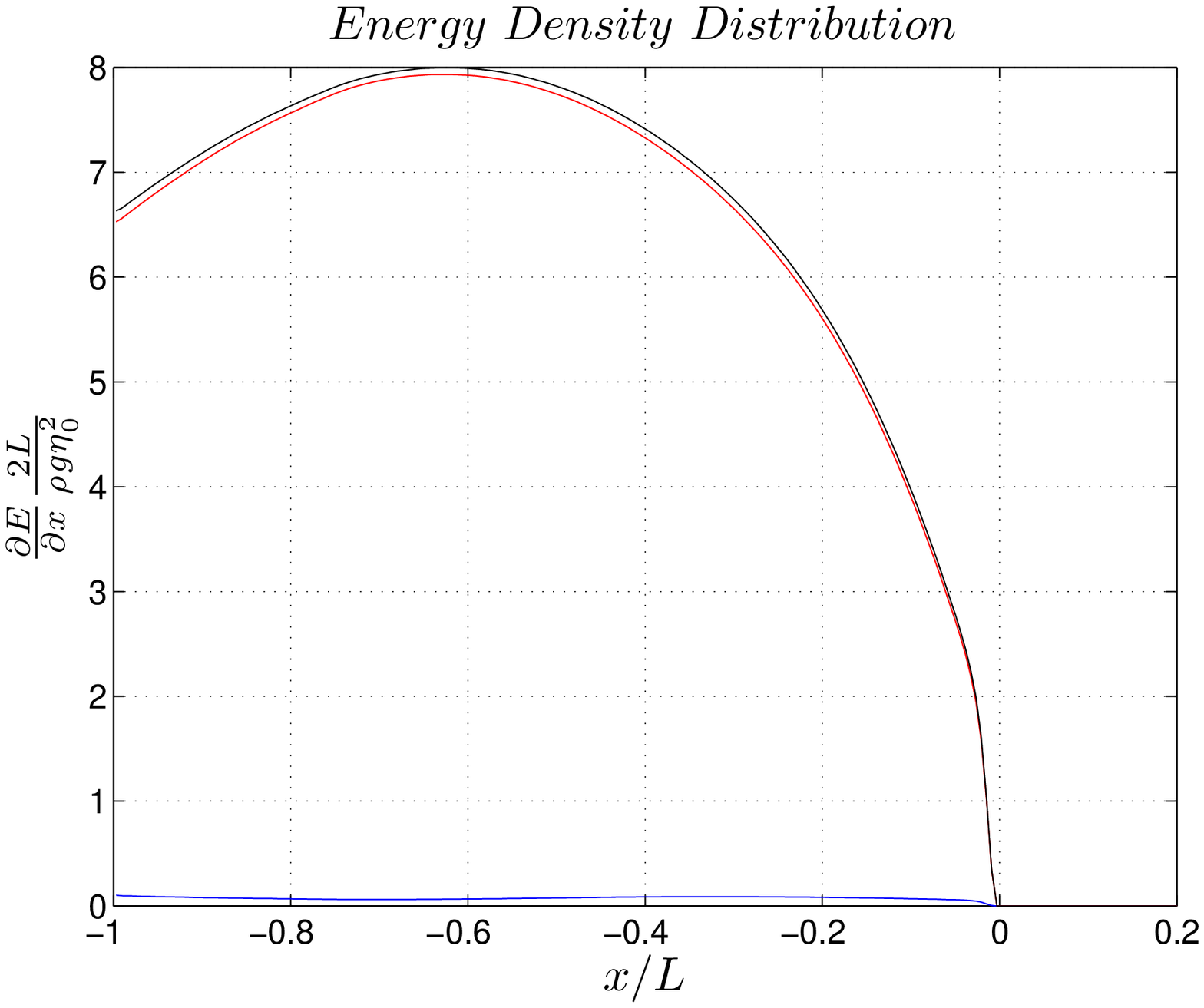}}
{\small (d)}
\end{minipage}
\caption{Energy density distribution during runup for the case of the resonant frequency ($\omega = 0.4\, $s$^{-1}$) when $\tan\theta = 0.13$ and $L = 12.5$ m, at the arrival of the first (a), second (b), third (c) and fourth (d) waves. Note: the scale of the vertical axis differs between the four snapshots and the color code is the same as in Fig. \ref{fig:sin_engy}.}
\label{fig:EngyDensity}
\end{center}
\end{figure}

In Fig. \ref{fig:EngyDensity}, four different snapshots of the energy density distribution during runup are shown in order to shed more light on the resonant mechanism ($\omega = 0.4\, $s$^{-1}\ ,\ \tan\theta = 0.13$). The first snapshot is taken at the instant when the first incident wave hits the initial shoreline. The potential energy is higher than the kinetic energy and both of them are concentrated close to the shoreline. After the runup of the first wave, the energy is reflected offshore, while at the same time energy is transferred shoreward by the second incident wave causing an amplification of kinetic energy. The same process is repeated by the following incident and reflected waves until the quasi-periodic state is reached. What is interesting is that the horizontal location where the amplification takes place remains almost stationary across runups and lies closer to the left boundary than the initial shoreline. After energy is amplified locally, it travels shorewards possibly due to the continuous forcing at the left boundary.

\begin{figure}
\begin{center}
\begin{minipage}{2.2in}
{\includegraphics[height=2.1in]{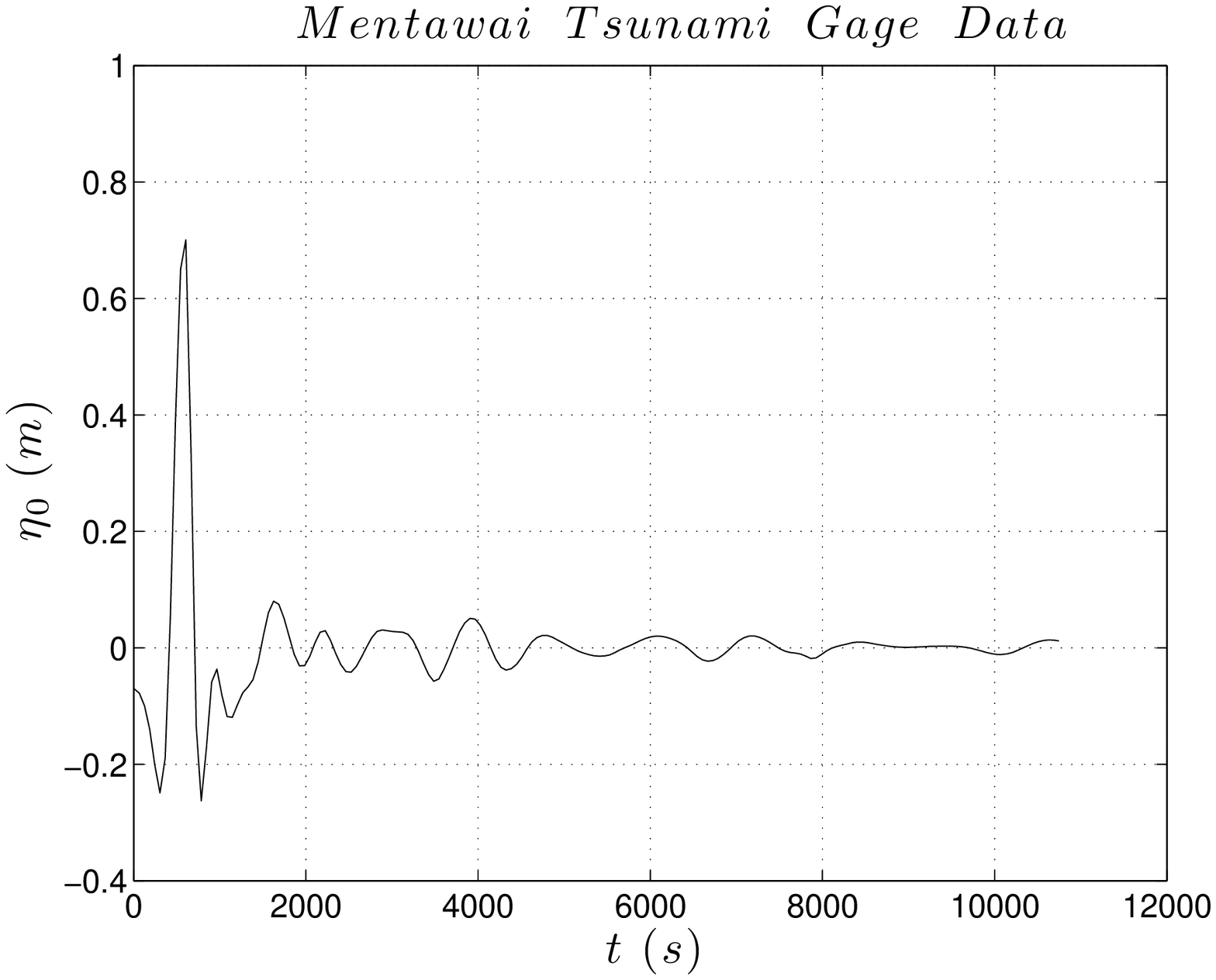}}
{\small (a)}
\end{minipage}
\hspace{0.2in}
\begin{minipage}{2.2in}
{\includegraphics[height=2.1in]{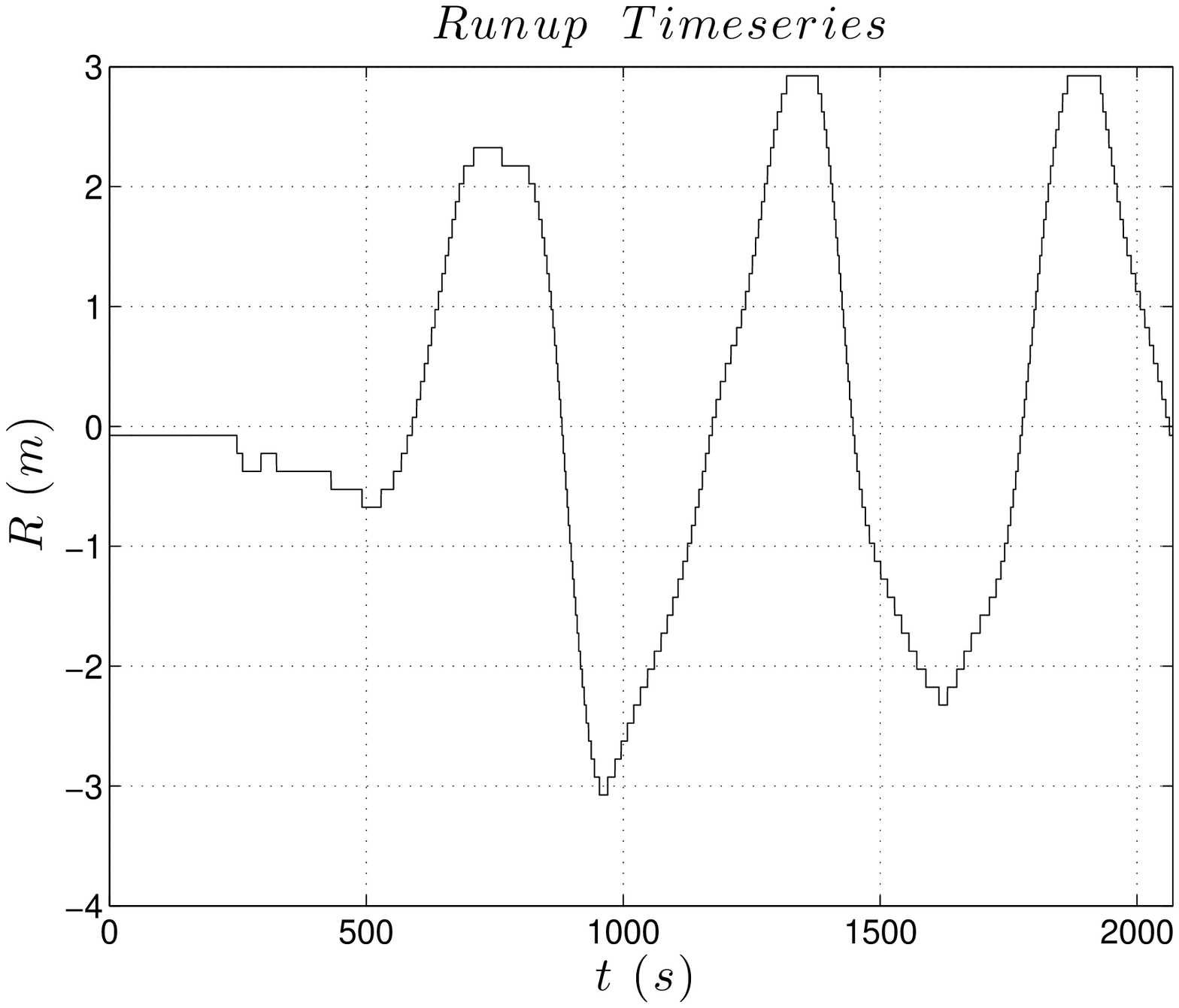}}
{\small (b)}
\end{minipage}
\caption{Virtual wave-gage (Lon = $100.24^o$ E\ , \  Lat = $-3.4^o$ N) data obtained for the 25 October 2010 Mentawai Islands tsunami (a). Timeseries of the shoreline elevation during the first 2000 s (b).}
\label{fig:GageData}
\end{center}
\end{figure}

Apart from idealistic simulations with sinusoidal waves, we explored whether similar resonant phenomena can occur during a real tsunami. Therefore a simulation was run for the 25 October 2010 Mentawai Islands tsunami. A virtual wave-gage was placed at Lon = $100.24^o$ E\ , \  Lat = $-3.4^o$ N, where the depth is approximately 120 m and the free surface elevation was obtained for the first $10800$ s of the tsunami (Fig. \ref{fig:GageData}\,a). From that data only the first $2000$ s were used as boundary value using a uniform slope $\tan\theta = 0.03$, which is close to the actual mean slope from the location of the wave-gage to the closest shore (the distance of the wave-gage to the shore is $L = 4000$ m). The timeseries of the shoreline elevation is shown in Fig. \ref{fig:GageData} (b). We can observe the runup of three waves at $t=720$ s, $t=1320$ s and $t=1860$ s. It is clear that the first wave does not cause the highest runup, even though it has the highest amplitude, as recorded by the wave-gage.

The fact that the highest runup is not driven by the leading and highest wave excited our curiosity to investigate whether there exists a connection between the resonant mechanism observed when using sinusoidal wave profiles and the wave-gage recordings. From Fig. \ref{fig:GageData} (b) one can see that the maximum runups are separated by approximately $600$ s intervals. If we assume that the incident wave is a sum of sinusoidal waves and $T = 600$ s is the period of the dominant mode, we can find the wavelength of that mode using $\lambda_0 = T \sqrt{gL\, \tan\theta}$. By doing so, the ratio $\lambda_0/L$ is equal to $5.15$  which according to our previous results (Fig. \ref{fig:sin_ampli}) corresponds to the resonant regime. Consequently, local resonant amplification of tsunami runup may explain why in some cases it is not the first wave that results in the highest runup.

\begin{figure}
\begin{center}
\begin{minipage}{2.2in}
{\includegraphics[height=2.0in]{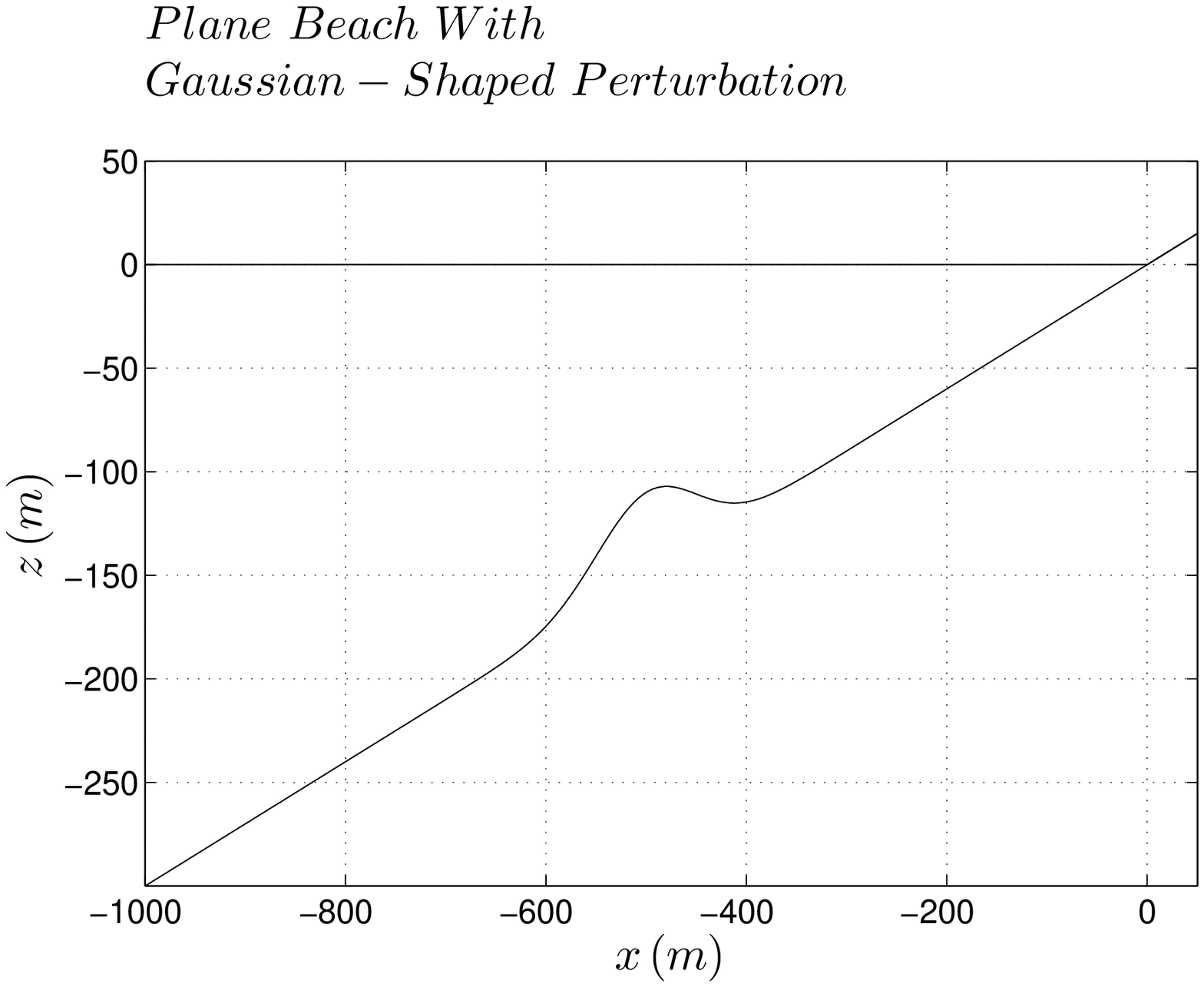}}
{\small (a)}
\end{minipage}
\hspace{0.2in}
\begin{minipage}{2.2in}
{\includegraphics[height=2.0in]{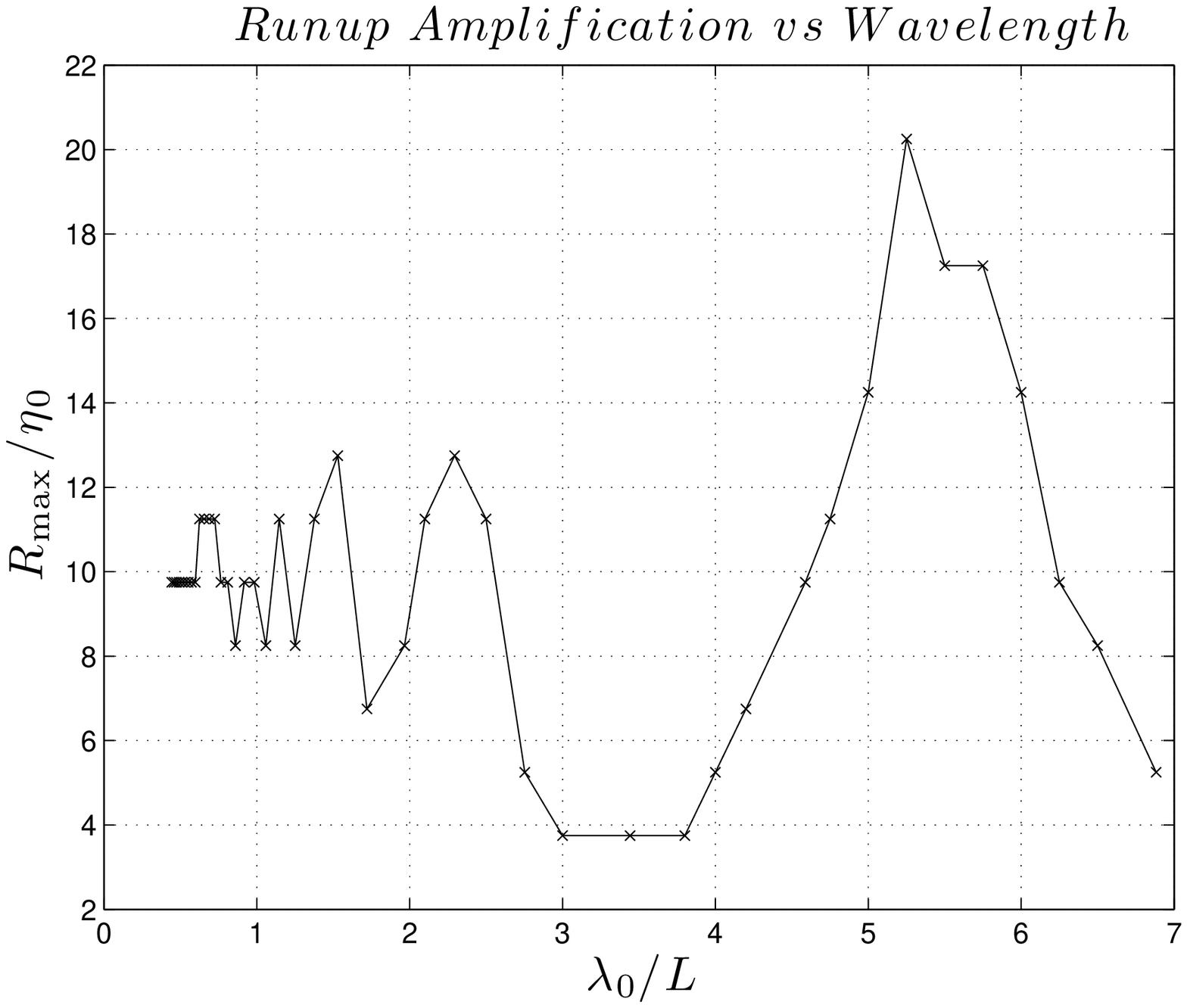}}
{\small (b)}
\end{minipage}
\begin{minipage}{2.2in}
{\includegraphics[height=2.0in]{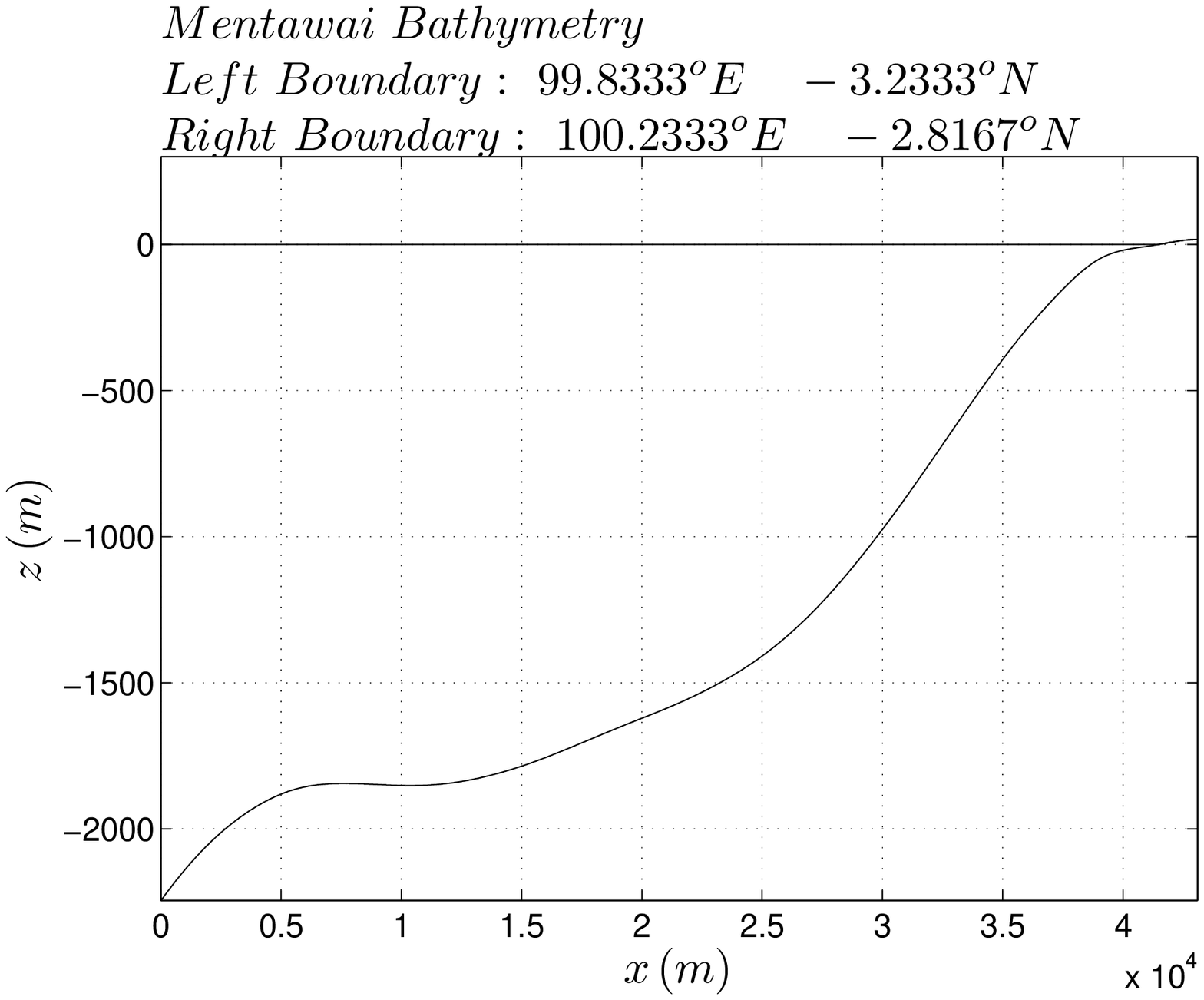}}
{\small (c)}
\end{minipage}
\hspace{0.2in}
\begin{minipage}{2.2in}
{\includegraphics[height=2.0in]{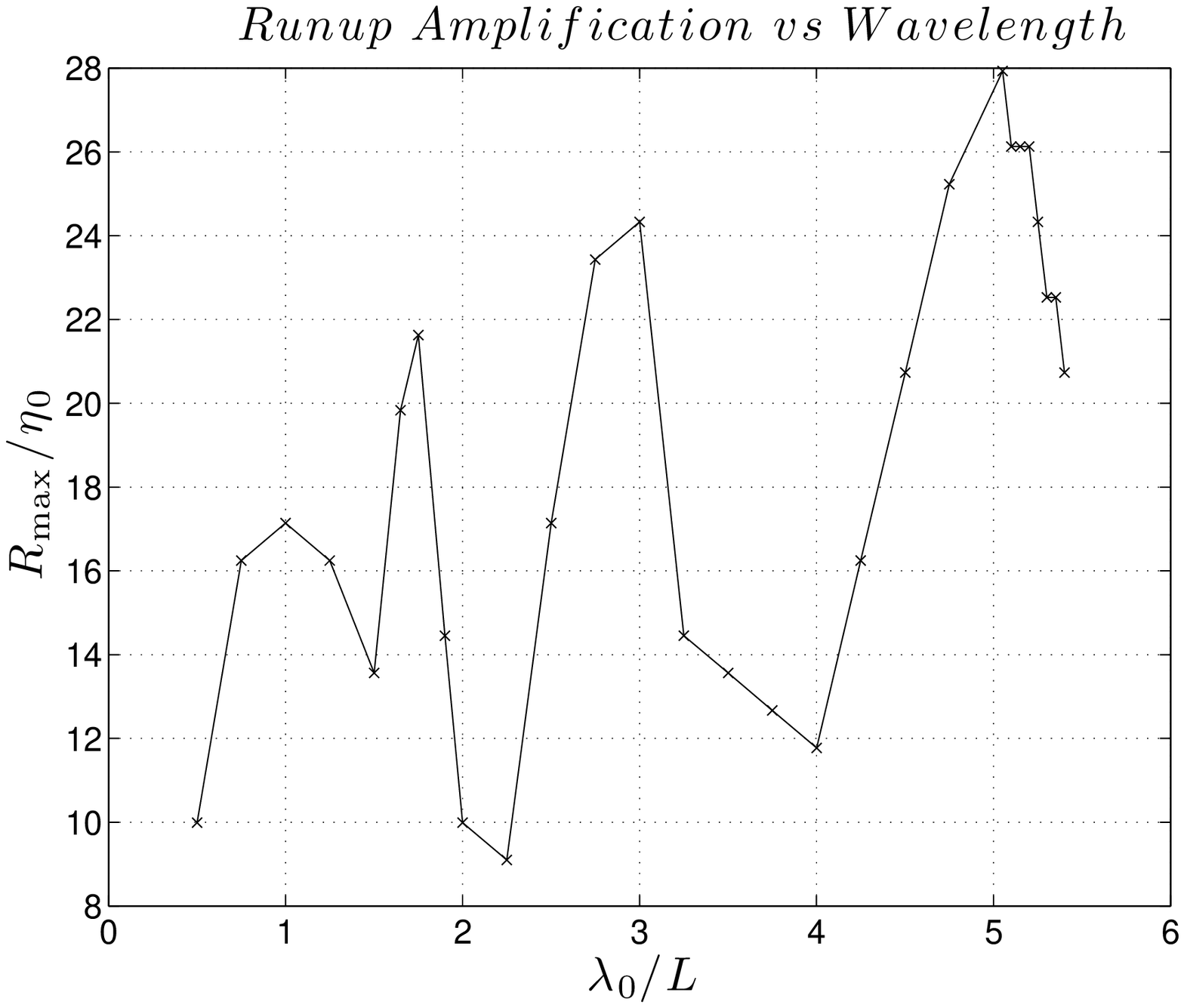}}
{\small (d)}
\end{minipage}
\caption{Plane beach perturbed by a Gaussian-shaped bathymetric feature (a). Amplification ratio as a function of non-dimensional wavelength (b). Bathymetry in the Mentawai Islands region (c). Amplification ratio as a function of non-dimensional wavelength (d).}
\label{fig:gauss}
\end{center}
\end{figure}

In addition to simulations with a plane beach we investigated two cases of non-trivial bathymetry. The first consists of a beach perturbed by a Gaussian-shaped underwater feature as in Fig. \ref{fig:gauss} (a). Again the forcing at the boundary was an idealized sinusoidal signal though this time it was limited to only four periods, since in nature one would not expect a wave-train larger than that. In Fig. \ref{fig:gauss} (b) we can observe the existence of resonant frequencies though now the amplification is not as high. What is intriguing is the existence of multiple peaks, signifying that resonant phenomena might occur much more often than expected. We reached the same conclusion when we studied the second case, which had a real bathymetry taken from the region of the Mentawai islands (Fig. \ref{fig:gauss}\,c). Multiple resonant frequencies can also be observed in this case (Fig. \ref{fig:gauss}\,d), thus further strengthening the suggestion that resonant runup amplification due to wave interactions is not a rare phenomenon.

\section{Conclusions}

In summary, we discovered local resonant amplification phenomena related to the one-dimensional BVP of the NSWE on a plane beach. The resonance occurs due to incoming and reflected wave interactions and the actual amplification ratio depends on the beach slope. These phenomena can explain why it is not always the first wave that causes the highest runup, as well as why the tail of a single wave may produce leading-order runup values. Resonant mechanisms are not limited to the plane beach paradigm but can be observed in more complex bathymetries as well, thus suggesting that local runup amplification is not a rare event. However, when the bathymetry is non-trivial, it is not clear to what extent resonance is attributed to wave trapping and generation of harmonics.

\section*{Acknowledgement}

We would like to thank Professors C. C. Mei and C. E. Synolakis for fruitful discussions. This work has been partially supported by the 2008 Framework Program for Research, Technological development and Innovation of the Cyprus Research Promotion Foundation under the Project A$\Sigma$TI/0308(BE)/05.

D.~Dutykh acknowledges the support from French Agence Nationale de la Recherche, project MathOcean (Grant ANR-08-BLAN-0301-01). F.~Dias and D.~Dutykh acknowledge the support of the Ulysses Program of the French Ministry of Foreign Affairs under the project 23725ZA.

\bibliography{biblio}

\newcommand{\etalchar}[1]{$^{#1}$}
\begin{thebibliography}{FKM{\etalchar{+}}07}

\bibitem[AB07]{Antuono2007}
M.~Antuono and M.~Brocchini.
\newblock The boundary value problem for the nonlinear shallow water equations.
\newblock {\em Stud. Appl. Math.}, 119:73--93(21), 2007.

\bibitem[AB10]{Antuono2010}
M.~Antuono and M.~Brocchini.
\newblock Solving the nonlinear shallow-water equations in physical space.
\newblock {\em J. Fluid Mech}, 643:207--232, 2010.

\bibitem[AM88]{Agnon1988}
Yehuda Agnon and Chiang~C. Mei.
\newblock Trapping and resonance of long shelf waves due to groups of short
  waves.
\newblock {\em J. Fluid Mech}, 195:201--221, 1988.

\bibitem[BP96]{Brocchini1996}
M.~Brocchini and D.~H. Peregrine.
\newblock Integral flow properties of the swash zone and averaging.
\newblock {\em J. Fluid Mech}, 317:241--273, 1996.

\bibitem[Car66]{Carrier1966}
G.~F. Carrier.
\newblock Gravity waves on water of variable depth.
\newblock {\em J. Fluid Mech}, 24(04):641--659, 1966.

\bibitem[CG58]{CG58}
G.~F. Carrier and H.~P. Greenspan.
\newblock Water waves of finite amplitude on a sloping beach.
\newblock {\em J. Fluid Mech.}, 2:97--109, 1958.

\bibitem[DD09]{Dutykh2009b}
D.~Dutykh and F.~Dias.
\newblock Energy of tsunami waves generated by bottom motion.
\newblock {\em Proc. R. Soc. A}, 465:725--744, 2009.

\bibitem[DKM11]{Dutykh2010}
D.~Dutykh, Th. Katsaounis, and D.~Mitsotakis.
\newblock Finite volume schemes for dispersive wave propagation and runup.
\newblock {\em Journal of Computational Physics}, 230:3035--3061, 2011.

\bibitem[DP08]{Didenkulova2008}
I.~Didenkulova and E.~Pelinovsky.
\newblock Run-up of long waves on a beach: the influence of the incident wave
  form.
\newblock {\em Oceanology}, 48(1):1--6, 2008.

\bibitem[FKM{\etalchar{+}}07]{Fritz2007}
H.~M. Fritz, W.~Kongko, A.~Moore, B.~McAdoo, J.~Goff, C.~Harbitz, B.~Uslu,
  N.~Kalligeris, D.~Suteja, K.~Kalsum, V.~V. Titov, A.~Gusman, H.~Latief,
  E.~Santoso, S.~Sujoko, D.~Djulkarnaen, H.~Sunendar, and C.~Synolakis.
\newblock Extreme runup from the 17 {J}uly 2006 {J}ava tsunami.
\newblock {\em Geophys. Res. Lett.}, 34:L12602, 2007.

\bibitem[GM03]{Grataloup2003}
G\'eraldine~L. Grataloup and Chiang~C. Mei.
\newblock Localization of harmonics generated in nonlinear shallow water waves.
\newblock {\em Phys. Rev. E}, 68(2):026314, Aug 2003.

\bibitem[Kaj77]{Kajiura1977}
Kinjiro Kajiura.
\newblock Local behaviour of tsunamis.
\newblock In D.~Provis and R.~Radok, editors, {\em Waves on Water of Variable
  Depth}, volume~64 of {\em Lecture Notes in Physics}, pages 72--79. Springer
  Berlin / Heidelberg, 1977.

\bibitem[KK64]{Keller1964}
J.B. Keller and H.B. Keller.
\newblock Water wave run-up on a beach.
\newblock Technical Report NONR-3828(00), Department of the Navy, Washington,
  DC, 1964.

\bibitem[KS98]{Kanoglu1998}
U.~Kanoglu and C.E. Synolakis.
\newblock Long wave runup on piecewise linear topographies.
\newblock {\em J. Fluid Mech.}, 374:1--28, 1998.

\bibitem[Mil67]{Miles1967}
John~W. Miles.
\newblock Surface-wave scattering matrix for a shelf.
\newblock {\em J. Fluid Mech}, 28(04):755--767, 1967.

\bibitem[MS10]{Madsen2010}
P.~A. Madsen and H.~A. Schaffer.
\newblock Analytical solutions for tsunami runup on a plane beach: single
  waves, n-waves and transient waves.
\newblock 645:27--57, 2010.

\bibitem[NSS{\etalchar{+}}11]{Neetu2011}
S.~Neetu, I.~Suresh, R.~Shankar, B.~Nagarajan, R.~Sharma, S.~Shenoi,
  A.~Unnikrishnan, and D.~Sundar.
\newblock Trapped waves of the 27 november 1945 makran tsunami: observations
  and numerical modeling.
\newblock {\em Nat. Hazards}, pages 1--10, 2011.

\bibitem[PM92]{Pelinovsky1992}
E.~N. Pelinovsky and R.~Kh. Mazova.
\newblock Exact analytical solutions of nonlinear problems of tsunami wave
  run-up on slopes with different profiles.
\newblock {\em Nat. Hazards}, 6:227--249, 1992.

\bibitem[Sor97]{Sorensen1997}
R.M. Sorensen.
\newblock {\em Basic coastal engineering}.
\newblock Springer, 1997.

\bibitem[Syn86]{Synolakis1986}
C.~Synolakis.
\newblock {\em The runup of long waves}.
\newblock PhD thesis, California Institute of Technology, 1986.

\bibitem[Syn87]{Synolakis1987}
C.~Synolakis.
\newblock The runup of solitary waves.
\newblock {\em J. Fluid Mech.}, 185:523--545, 1987.

\bibitem[TS94]{TS94}
S.~Tadepalli and C.~E. Synolakis.
\newblock The run-up of {N}-waves on sloping beaches.
\newblock {\em Proc. R. Soc. Lond. A}, 445:99--112, 1994.

\end{thebibliography}
\bibliographystyle{alpha}

\end{document}